\documentclass{elsart5}

\usepackage{graphicx}

\usepackage{amssymb}
\usepackage{amsmath}

\begin{document}

\begin{frontmatter}

\title{Enhanced low-temperature entropy and flat-band ferromagnetism in
the $t-J$ model on the sawtooth lattice}

\author[aff1]{A. Honecker\corauthref{cor1}\thanksref{thanks1}}
\ead{ahoneck@uni-goettingen.de}
\thanks[thanks1]{Participation at ICM2006 supported by the
DFG through SFB 602.}
\corauth[cor1]{}
\author[aff2]{J. Richter}
\address[aff1]{Institut f\"ur Theoretische Physik,
Universit\"at G\"ottingen,
Friedrich-Hund-Platz 1,
37077 G\"ottingen,
Germany}
\address[aff2]{Institut f\"ur Theoretische Physik, Otto-von-Guericke
 Universit\"at Magdeburg,
 39016 Magdeburg, Germany}
\received{7 June 2006} 
\volume{310}
\issue{2}
\pubyear{2007}
\copyear{2006}
\firstpage{1331}
\lastpage{1333}



\begin{abstract}
Using the example of the sawtooth chain, we argue that the $t-J$ model
shares important features with the Hubbard model on highly
frustrated lattices. The lowest single-fermion band is completely
flat (for a specific choice of the hopping parameters $t_{i,j}$ in the case
of the sawtooth chain), giving rise to single-particle
excitations which can be localized in real space. These localized
excitations do not interact for sufficient spatial separations such
that exact many-electron states can also be constructed.
Furthermore, all these excitations acquire zero energy for a suitable choice
of the chemical potential $\mu$. This leads to:
(i) a jump in the particle density at zero temperature,
(ii) a finite zero-temperature entropy,
(iii) a ferromagnetic ground state with a charge gap when the flat
    band is fully occupied and
(iv) unusually large temperature
variations when $\mu$ is varied adiabatically at finite temperature.
\end{abstract}


\end{frontmatter}

During the past years, it has been noted that exact ground states
can be constructed for the antiferromagnetic $XXZ$ model at high fields
on a large class of highly frustrated lattices
(see \cite{jump,RSH04,RSHSS04} and references therein).
This leads to a finite zero-temperature entropy exactly at the saturation
field and an enhanced magnetocaloric effect
\cite{RSH04,ZhiHo,DeRi,ZhiTsu}, suggesting potential applications for
efficient low-temperature magnetic refrigeration \cite{ZhiHo,Zhito}.
Recently, we have pointed out \cite{HoRi} analogies to
flat-band ferromagnetism in the Hubbard model on the same
lattices (see e.g.\ \cite{Mielke92,Tasaki92,Tasaki98,IKWO98,NGK03}).

\begin{figure}[t]
\begin{center}
\includegraphics[width=0.96\columnwidth]{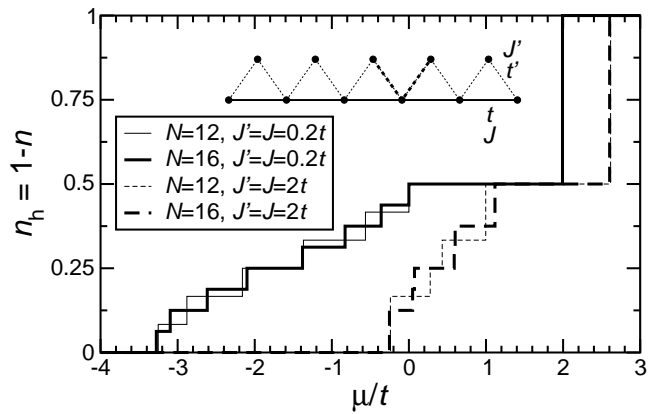}
\end{center}
\caption{Inset: The sawtooth chain model. Filled circles show
electron sites. The hopping (magnetic exchange) are $t$ ($J$)
along the base line and $t'$ ($J'$) along the dashed zigzag-line,
respectively. The bold valley shows the area occupied by a localized
excitation.
Main panel: Hole density $n_{\rm h} = 1-n$ at temperature $T=0$
as a function of chemical potential $\mu$ for $t' = \sqrt{2}\, t > 0$
and two choices of $J'=J$.
}
\label{nmu-fig}
\end{figure}

Here we will illustrate some of the issues with exact diagonalization
results for the $t-J$ model. The $t-J$ model
arises as the large-$U$ limit of the Hubbard model and
is defined by the Hamiltonian
\begin{eqnarray}
H &=& \sum_{\sigma} \sum_{\langle i, j\rangle} t_{i,j} \, P\, \left(
{c}^{\dagger}_{i,\sigma} {c}_{j,\sigma}
 + {c}^{\dagger}_{j,\sigma} {c}_{i,\sigma}\right)\,P \nonumber \\
&& + \sum_{\langle i, j\rangle} J_{i,j}
 \left(\vec{S}_i \cdot \vec{S}_j - {1 \over 4} \, n_i \, n_j \right)
+ \mu \sum_{i=1}^N {n}_{i} \, .
\label{HtJ}
\end{eqnarray}
The sums run over the nearest-neighbor pairs
$\langle i, j\rangle$ of a lattice with $N$
sites. ${c}^{\dagger}_{i,\sigma}$ and ${c}_{i,\sigma}$ are
the usual fermion creation and annihilation operators,
$P$ is the projector which eliminates doubly occupied sites,
${n}_i = {c}^{\dagger}_{i,\uparrow} {c}_{i,\uparrow} +
{c}^{\dagger}_{i,\downarrow} {c}_{i,\downarrow}$ is the
total number operator at site $i$, and $\vec{S}_i$ are spin-1/2
operators acting on an occupied site $i$.

Here we will concentrate on the sawtooth chain
model sketched in the inset of Fig.~\ref{nmu-fig}.
The lower of the two branches of the single-electron dispersion 
becomes completely flat for $t' = \sqrt{2}\,t$.
For this choice one can construct first
localized single-electron excitations living in one of the
valleys of the sawtooth chain (bold dashed line in the
inset of Fig.~\ref{nmu-fig}), and then
excitations with $N_{\rm el}$ electrons
which are non-interacting for sufficient spatial
separations and thus have energy $E = (-2\,t + \mu) \, N_{\rm el}$,
in exactly the same manner as for the
Hubbard model \cite{HoRi}. So far, the magnetic exchanges
$J_{i,j}$ are arbitrary. However, it will turn
out that they should be chosen sufficiently weak in order
to ensure that the non-interacting localized many-electron
states are the ground states in their respective particle number
subspaces.
At half filling $n = \langle n_i \rangle = 1$ only the magnetic
part of the $t-J$ model survives such that it reduces to the
previously studied antiferromagnetic spin-1/2 Heisenberg model
(see \cite{jump,RSHSS04,ZhiHo,DeRi,ZhiTsu} for the sawtooth chain).

The main panel of Fig.~\ref{nmu-fig} shows finite-system results for
$n_{\rm h} = 1-n$ at $T=0$ versus $\mu$ for $t' = \sqrt{2}\,t$
(these curves are the electronic counterpart of the magnetization curves
\cite{RSHSS04}). For small magnetic exchange (like $J' = J = 0.2\,t$),
there is a jump of height $\delta n = 1/2$ exactly at $\mu = 2\,t$.
At this point, all localized many-electron excitations
collapse to $E=0$.
Furthermore, for $N=12$ the number of ground states
is 1, 12, 54, 112, 105, 36, 7 in the sectors with $N_{\rm el} = 0$, 1, 2, 3,
4, 5, 6, respectively. This leads to a ground-state entropy per site
$\ln(327)/12 = 0.48\ldots$ at $\mu = 2\,t$ for $N=12$. The ground-state
degeneracies are exactly the same as for the Hubbard model \cite{hubbard}
consistent with the ground states of the $t-J$ model for small
$J_{i,j}$ and $N_{\rm el} \le N/2$ being projections of those
of the Hubbard model. General theorems for the Hubbard model
imply a saturated ferromagnetic ground state
for $N_{\rm el} = N/2$ (see e.g.\ \cite{Tasaki98,IKWO98,NGK03} for
the sawtooth chain).
Numerically, we find a fully saturated ferromagnet for the
$t-J$ model in the sectors with $N_{\rm el} = N/2$
and $N/2-1$. The plateau at $n=N_{\rm el}/N=1/2$ in
the $n(\mu)$-curve in Fig.~\ref{nmu-fig} shows that
the ground state is a saturated ferromagnet for $0 < \mu < 2\,t$,
corresponding to an appreciable charge gap.

The situation changes for larger antiferromagnetic
$J_{i,j}$, as illustrated for $J'=J = 2 \,t$ in Fig.~\ref{nmu-fig}.
In this case the localized states are no longer the
lowest-energy states. This is signalled by a shift of the jump between
$n=1/2$ and $n=0$ to $\mu > 2\,t$ which now corresponds to a
true first-order transition. The charge gap, {\it i.e.},
the plateau at $n=1/2$ is also present in this case.

\begin{figure}[t]
\begin{center}
\includegraphics[width=0.96\columnwidth]{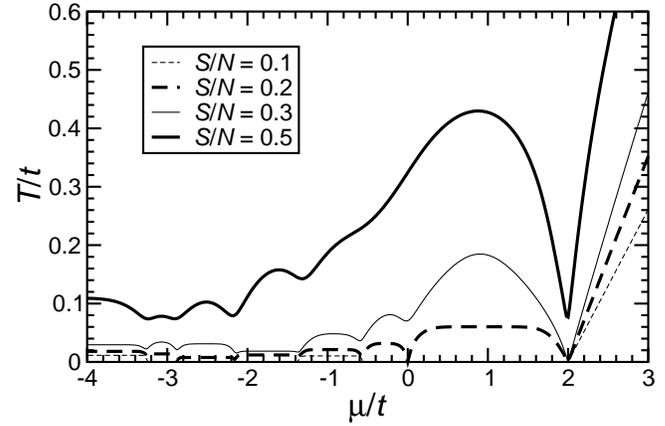}
\end{center}
\caption{Curves of constant entropy $S$ for $t' = \sqrt{2}\, t > 0$,
$J'=J = 0.2\, t$, and $N=12$ sites.
}
\label{constS-fig}
\end{figure}

The ground-state degeneracies are reflected by thermodynamic
properties, as illustrated for the entropy $S$ in Fig.~\ref{constS-fig}
(the curves of constant $S$ correspond to the adiabatic demagnetization
curves of the magnetic counterpart \cite{ZhiHo}). In particular,
the finite $T=0$ entropy at $\mu = 2\,t$ leads to large temperature
changes during adiabatic variations of $\mu$, even
cooling to $T=0$ as $\mu \to 2\,t$ at low temperatures. The low-temperature
properties for $\mu$ close to $2\,t$ are controlled by the localized
states and are independent of the details of the microscopic model
($J_{i,j}$ in the $t-J$ model and
$U$ in the Hubbard model \cite{hubbard}); finite-size effects
are also small in this region. By contrast, the behavior for $\mu < 0$
in Fig.~\ref{constS-fig} exhibits strong finite-size effects at low
temperatures and depends on details of the model: for example, in this
region the presence of doubly occupied sites leads to qualitatively
different behavior of the Hubbard model \cite{hubbard}.

We have focussed on the sawtooth chain, but it should
share important features with
a large class of highly frustrated lattices such as
the kagom\'e lattice \cite{jump,ZhiTsu,Mielke92} which do
not require any fine-tuning. We expect that
the $t-J$ model with weak $J_{i,j}$ has the same localized
excitations as the repulsive Hubbard model such that it shares
in particular the same properties with respect to flat-band
ferromagnetism \cite{Mielke92,Tasaki92,Tasaki98,IKWO98,NGK03}.
The main advantage of the $t-J$ model is a substantially reduced
Hilbert space dimension close to $n=1$
which simplifies a full diagonalization and
thus the exact determination of finite-temperature properties
of a finite system.

\end{document}